\documentstyle[aps]{revtex}
\begin{document}

\def\half{{1 \over 2}}
\newcommand{\be}{\begin{equation}}
\newcommand{\ee}{\end{equation}}
\newcommand{\bea}{\begin{eqnarray}}
\newcommand{\eea}{\end{eqnarray}}

\twocolumn[\hsize\textwidth\columnwidth\hsize\csname
@twocolumnfalse\endcsname

\title{Yes, semiclassical zero temperature black holes exist!}

\author{David A. Lowe}

\address{Department of Physics, Brown University, Providence, RI
  02912}

\maketitle
\vskip2pc]

In a recent letter \cite{anderson}, the claim was made that ``in all
physically realistic cases, macroscopic zero temperature black hole
solutions do not exist.'' We will show this conclusion was reached on
the basis of an incorrect calculation.

The Reissner-Nordstr\"om metric is parameterized as
\bea
\label{metric}
ds^2 &= &-( 1 + 2 \epsilon \rho(r)) (1 - {2 m(r) \over r} +{Q^2 \over
  r^2}) dt^2  \nonumber \\
& +& (1-{2 m(r) \over r} + {Q^2\over r^2})^{-1} dr^2 + r^2 d 
\Omega^2~.
\eea
Defining $m(r) = M(1+\epsilon \mu(r))$, with $\epsilon = \hbar/M^2$, 
the authors find the semiclassical Einstein equations
\bea
\label{einstein}
{d \mu \over dr} &=& - {4 \pi r^2 \over M \epsilon} \langle
T^t_t\rangle \nonumber \\ 
{d \rho \over dr} &=& {4 \pi r \over \epsilon} (1 - {2 M \over r} +{Q^2 
  \over r^2} )^{-1}( \langle T^r_r\rangle -\langle T^t_t\rangle)
\nonumber \\
\eea
and try to find a perturbative solution for $\mu$ and $\rho$. At the
unperturbed horizon $r_+ = M+\sqrt{M^2-Q^2}$, 
the authors set $\mu(r_+)= C_1$ and $\rho(r_+)=C_2$.

They then go on to state that
the perturbed horizon lies at $r=M_R + \sqrt{M_R^2-Q^2}$ where $M_R= M(1+
\epsilon C_1)$. The perturbed horizon is a solution of 
\be
\label{horpos}
r^2 - 2 m(r) r +Q^2 =0~.
\ee
Here they perform a double expansion in $\epsilon$ and $r-r_+$,
keeping $|r-r_+| \leq {\cal O}(\epsilon)$.
Keeping only terms of order $\epsilon$ in $m$ gives the result the
authors find. However when one approaches the extremal limit ($M^2-Q^2 
\leq {\cal O}(\epsilon)$), it becomes necessary to expand $m$ to order
$\epsilon^2$ to obtain $r_h$ correctly to order $\epsilon$. 
This follows simply from the form of the solution to the quadratic
equation. When one does this the perturbed horizon does not lie at 
$M_R + \sqrt{M_R^2-Q^2}$. 

The correct answer at extremality  (expanding in $\epsilon$ and $r-M$, 
keeping $|r-M| \leq {\cal O}(\epsilon)$ and
denoting $d \mu/dr$ by $\mu'$) is 
\be
\label{correct}
m(r) = M - M^3 \epsilon^2 (\mu'(M))^2 + M \epsilon \mu'(M) (r-M) +\cdots
\ee
with the extremal charge $Q=M- \half \epsilon^2 M^3 (\mu'(M))^2+\cdots$.
Note $\mu'(M)$ is determined by (\ref{einstein}). The $r-M$
independent term in (\ref{correct}) is an integration constant of
(\ref{einstein}) which may be chosen for convenience. The quantum
correction to the extremality relation
between $M$ and $Q$ is fixed by demanding that the discriminant of the 
quadratic equation (\ref{horpos}) vanishes. The solution of
(\ref{horpos}) gives the position of the horizon
 $r_h= M + \mu'(M) \epsilon M^2 + \cdots$, to leading order
in $\epsilon$.
Inserting (\ref{correct}) into the
metric, one finds the surface gravity of the black hole vanishes on
the horizon. This is most easily seen by noting that 
(\ref{horpos})
has a double zero at $r=r_h$ (at order $\epsilon^2$).
This solution holds regardless of the sign of $\mu'(M)$, and smoothly
matches onto the classical solution as $\epsilon \to 0$, contradicting 
the 
calculation performed in \cite{anderson}.

I thank Don Marolf for helpful comments.

\end{document}